\documentclass[epj]{svjour}
\usepackage{epsfig}
\usepackage{color}
\usepackage{float}
\usepackage{amssymb}
\usepackage{amsmath, bbm}
\usepackage[figuresright]{rotating}

\usepackage[numbers,sort&compress]{natbib}
\usepackage{hyperref}

\newlength{\feynwidth} 

\newcommand{\La}{{\Lambda}}
\newcommand{\Si}{{\Sigma}}

\newcommand{\be}{\begin{eqnarray}}
\newcommand{\ee}{\end{eqnarray}}

\usepackage{graphics}
\usepackage{psfrag}
\usepackage{multirow,mathtools}
\usepackage{bbm}
\def\bea{\begin{eqnarray}}
\def\eea{\end{eqnarray}}
\def\beq{\begin{equation}}
\def\eeq{\end{equation}}
\def\nn{\nonumber}
\def\bp{\mathbf{p}}
\def\bk{\mathbf{k}}

\def\br{\mathbf{r}}

\def\bP{\mathbf{P}}
\newcommand{\Lc}{{\Lambda}_c}
\newcommand{\Sc}{{\Sigma}_c}

\begin{document}

\title{Femtoscopic correlations and the $\Lambda_c N$ interaction}
\titlerunning{$\Lambda_c N$ correlations }        % if too long for running head

\author{J. Haidenbauer$^1$, G. Krein$^2$, and T.~C. Peixoto$^{2,3}$}
\authorrunning{J. Haidenbauer, G. Krein, and T.~C. Peixoto}

\institute{
$^1$Institute for Advanced Simulation, Institut f\"{u}r Kernphysik (Theorie) and J\"{u}lich
Center for Hadron Physics, Forschungszentrum J\"{u}lich, D-52425 J\"{u}lich, Germany \\
$^2$Instituto de F\'{\i}sica Te\'orica, Universidade Estadual Paulista, 
Rua  Dr. Bento Teobaldo Ferraz, 271 - Bloco II, 01409-001 S\~ao Paulo, SP, Brazil \\
$^3$Instituto Federal de Educa\c{c}\~ao, Ci\^encia e Tecnologia de Sergipe,
Rodovia Juscelino Kubitschek, s/n, 49680-000 Nossa Senhora da Gl\'oria,
SE, Brazil
}
\date{Received: date / Accepted: date}

\abstract{
We study the prospects for deducing constraints on the interaction of charmed 
baryons with nucleons from measurements of two-particle momentum correlation 
functions for $\Lambda_c p$. The correlation functions are calculated for 
$\Lambda_c N$ and $\Sigma_c N$ interactions that have been extrapolated 
from lattice QCD simulations at unphysical masses of $m_\pi=410-570$~MeV 
to the physical point using chiral effective field theory as guideline. 
In addition, we consider phenomenological $Y_c N$ models from the 
literature to explore the sensitivity of the results to the properties 
of the interaction in detail. We find that a measurement of the 
$\Lambda_c p$ correlation functions could indeed allow one to 
discriminate  between strongly attractive $\Lambda_c N$ forces, as 
predicted by some phenomenological models, and a weakly attractive 
interaction as suggested by the presently available lattice simulations.
}

\maketitle

% - - - - - - - - - - - - - - - - - - - - - - - - - - - - - - 

\section{Introduction}
\label{sec:intro}

Two-hadron momentum correlation functions extracted from relativistic heavy-ion 
collisions provide a doorway to information on the hadron-hadron interaction at 
low energies~\cite{Koonin:1977fh,Lednicky:1981su}, presently inaccessible 
by other means. This concerns especially the interaction of charmed
hadrons with ordinary matter, for example the one of the charmed 
baryons $\Lc$ and $\Sc$ ($Y_c$) with nucleons. 
Insight into the dynamics of such systems would deepen our notion 
of the flavor dependence of the strong interaction,  encoded on the fundamental
level in quantum chromodynamics~(QCD). Indeed,
the understanding of the flavor dependence of hadron-hadron 
forces is a key element in the study of charmed dibaryons~\cite{Karliner:2015ina}
and exotic hadronic molecules~\cite{Guo:2017}. The lack of knowledge on the 
$Y_cN$ interaction also hinders progress in the long-standing issue regarding the
existence of charmed nuclei~\cite{Tyapkin,Dover:1977jw,
Iwao:1976yi,Gatto:1978ka,Bhamathi:1981yu,Kolesnikov:1981nd,Bando:1981ti,Bando:1983yt,
Gibson:1983zw,Bando:1985up,Bhamathi:1989yb,Bunyatov:1992in,Tsushima:2002ua,
Tsushima:2003dd,Kopeliovich:2007kd,Miyamoto:2017tjs,Vidana:2019amb,
Haidenbauer:2020uci}{\textemdash}for recent reviews see~\cite{Hosaka:2016ypm,
Krein:2017usp,Krein:2019gcm}. These are nuclei containing a $Y_c$ hyperon,
similar to the more familiar hypernuclei which are formed with a strange baryon, 
$\Lambda$ and/or $\Sigma$ ($Y$). 

The discovery of $Y_c$ hypernuclei would reveal a new form of strongly-interacting matter and 
thereby widen our knowledge on the QCD phase diagram. It would also give hope to learn about 
medium effects on the phenomenon of chiral symmetry restauration, a phenomenon 
associated with the light quarks and sensitive to environmental effects, which 
a $Y_c$ would probe when bound to a nucleus~\cite{Carames:2018xek}. 
Although, in principle, dedicated scattering 
experiments producing low-energy $Y_c$ hyperons might be feasible in the future at sites 
such as J-PARC~\cite{Noumi:2017hbc} and KEK~\cite{Niiyama:2017wpp} in Japan and
FAIR~\cite{Friman:2011zz,Wiedner:2011} in Germany, high-energy heavy-ion 
experiments produce enough $Y_c$ hyperons (and nucleons, of course) to facilitate the
extraction of a $Y_cN$ momentum correlation function. Given the prospects and no 
impediment of principle, this is a timely opportunity worth exploring.

Actually, the opportunity offered by heavy-ion collisions and/or high-energetic $pp$ 
collisions has been already successfully exploited in respective investigations of the 
$\La p$, $\Si^0 p$, and $\Xi^- p$ 
systems~\cite{Adams:2005,Agakishiev:2010qe,Adamczewski-Musch:2016jlh,Acharya:2018gyz,Acharya:2019kqn,Acharya:2019sms}. 
Femtoscopic studies of the $YN$ interaction certainly profit from the large $\Lambda$ production yields, 
which are much larger than those of $\Lambda_c$. Yet, recent $pp$, $pA$ and $AA$ 
experiments~\cite{Zhou:2017ikn,Acharya:2017kfy,Acharya:2018ckj,Adam:2019hpq,
Meninno:2019jug,Sirunyan:2019fnc,Vermunt:2019ecg} discovered far greater $\Lambda_c$ 
yields than predicted by traditional hadronization models, which 
is welcome news for extracting a $Y_cN$ correlation function from such collisions. 
From the theoretical side, the $\Lambda_c N$ system benefits from the
absence of nearby thresholds, the presence of which would require a 
coupled-channels approach and would also introduce further 
uncertainties~\cite{Lednicky:1998,Haidenbauer:2018jvl}.
Indeed, in case of $\Lc N$ the nearest other threshold ($\Sc N$) 
is separated by an energy of $M_{\Sc} - M_{\Lc} = 168$~MeV, 
whereas for~$YN$ and the $\La N$ system it is separated by 
just $M_\Sigma - M_\Lambda = 78$~MeV. These positive perspectives motivate us to 
utilize the available theoretical information on the $Y_c N$ force to predict $\Lc N$
momentum correlation functions with the aim to initiate pertinent femtoscopic experiments.

Most of the theoretical work on the $Y_cN$ force has been done within meson-exchange 
models. Refs.~\cite{Vidana:2019amb,Liu:2011xc} are the most recent examples. 
There is also the very recent quark-model based study of 
Ref.~\cite{Garcilazo:2019ryw},
and that of Ref.~\cite{Maeda:2015hxa}, which combines both models. Although not 
constrained by experimental data, some of the studies do rest on symmetry 
principles and physical consistency. In meson-exchange models, SU(4) flavor symmetry, 
albeit questionable in the charm sector~\cite{Navarra:1998vi,Fontoura:2017ujf}, 
constrains the values of coupling constants. In quark models, fitting the low-lying
hadron spectrum and hadron-hadron scattering obvervables constrains parameters 
such as quark masses and quark-quark forces. 
Nonetheless, the overall situation is certainly unsatisfactory. 
However, it started to change with the recent lattice QCD (LQCD) simulations 
by the HAL QCD Collaboration~\cite{Miyamoto:2017,Miyamoto:2017ynx}. 
The HAL QCD results are for unphysical quarks masses, corresponding
to $m_\pi=410$ MeV or larger, and thus, need to be extrapolated to the 
physical point if one wants to see the proper physical implications. 
Ref.~\cite{Haidenbauer:2017dua} carried out such an extrapolation with chiral 
effective field theory (EFT) techniques~\cite{Epelbaum:2002gb,Petschauer:2013uua}, 
following the scheme of Refs.~\cite{Polinder:2006zh,Haidenbauer:2013oca,
Haidenbauer:2019boi} used for the $YN$ system. The overall theoretical picture 
revealed by the phenomenological and lattice studies can be summarized as 
follows: 1) the $\Lc N$ and $\Sc N$ forces are attractive, 2) the 
$\Lc N$ interaction from LQCD and its extrapolation to the physical point are 
much weaker than those suggested by most phenomenological studies. Given this
situation, there arises the question whether measurements of the $\Lc N$ correlation 
functions could allow one to discriminate between the model results
and the predictions based on/inferred from lattice simulations. 
In this paper, we give an affirmative answer to this question. 

The paper is organized as follows. In the next section, we provide a brief
overview of the formalism for evaluating two-hadron momentum correlation
functions.
In Sec.~\ref{sec:res} we introduce the employed $\Lc N$ interactions and
we provide predictions for the corresponding $\Lc p$ correlation 
functions. 
The paper closes with a Summary. 

%We conclude with a look at the prospects and challenges of extracting the 
%predicted correlation functions in heavy-ion experiments. 

% % % % % % % % % % % % % % % % % % % % % % % % % % % % % % % % % % % % % % % % % 
%
\section{Correlation function}
\label{sec:correl}

We summarize the main steps and compile the basic equations of femtoscopy to 
access hadron-hadron scattering information~\cite{Heinz:1999rw,Lisa:2005dd}. 
The extracted observable is a correlation function $C(\bp_1,\bp_2)$ 
of measured hadron momenta $\bp_1$ and $\bp_2$. $C(\bp_1,\bp_2)$ entails 
a ratio of two yields: $C(\bp_1,\bp_2) = A(\bp_1,\bp_2)/B(\bp_1,\bp_2)$, 
with $A(\bp_1,\bp_2)$ formed by hadrons coming from the same collision (coincidence yield) 
and $B(\bp_1,\bp_2)$ formed by hadrons coming from separate events (uncorrelated yield). 
A $C(\bp_1,\bp_2)$ not equal to unity implies correlation between the detected
particles; a correlation occurs due to mutual interaction and also due to quantum 
interference. The latter arises only for identical particles and, accordingly, is not present 
in the combination $Y_c$ and $N$. 

Experimental data on $C(\bp_1,\bp_2)$ and their theoretical interpretation 
are normally discussed in terms of the center-of-mass and relative momentum 
coordinates, $\bP = \bp_1 + \bp_2$ and $\bk = (M_2 \bp_1 - M_1 \bp_2)/(M_1+M_2)$, 
where $M_1$ and $M_2$ are the hadron masses. In terms of these coordinates, a connection 
between the measured correlation function and hadron-hadron scattering can be made in 
the  rest frame of the pair, $\bP=0$, through the (approximately valid) 
Koonin-Pratt formula~\cite{Koonin:1977fh,Pratt:1984su}:
\begin{equation}
C(\bk) = \frac{A(\bk)}{B(\bk)} 
\approx \int d\br \, S_{12} (\br)\, |\psi(\br,\bk)|^2 \ .
\label{KP}
\end{equation}
Here $\psi(\br,\bk)$ is the relative wave function of the pair and $S_{12}(\br)$ 
a static source distribution, a relative distance distribution in the pair's 
rest frame{\textemdash}Refs.~\cite{Heinz:1999rw,Lisa:2005dd,Bauer:1993wq,Anchishkin:1997tb} 
discuss the validity of the assumptions and approximations behind this formula. 

We compute the wave function $\psi(\br,\bk)$ within the formalism described in 
Ref.~\cite{Haidenbauer:2018jvl}. To~make the paper self-contained, we describe
the main features of that formalism but present only those equations relevant 
for this study. As we elaborate in the next section, coupled channels do not 
play
an important role, basically for the reason discussed in the Introduction. Therefore,
we restrict the formal part to the single-channel case~\cite{Haidenbauer:2018jvl}.

Past studies have shown that the correlations are predominantly due to the 
interaction in the $S-$waves. Accordingly, only the pertinent modifications in the 
$S-$wave part of the wave function, $\psi_{l=0}(r,k)  = \psi_0(r,k)$, 
are taken into account so that one can write~\cite{Lednicky:1981su,Ohnishi:2016elb}:
\begin{equation}
\psi(\br,\bk) = e^{i \bk\cdot\br} + \psi_0(r,k) - j_0(kr) \ ,   
\label{Eq:wfa}
\end{equation}
where $j_0(kr)$ is the $S-$wave component of the non-in\-ter\-acting wave function, 
a spherical Bessel function. Supposing a spherically symmetric source
$S_{12}(r)$, one obtains for the Koonin-Pratt formula: 
\begin{equation}
\hspace{-0.10cm}
C(k) = 1 + 4\pi \int dr \, r^2 \, S_{12}(r) \left[|\psi_0(k,r)|^2
- |j_0(kr)|^2 \right] \ .
\label{Eq:cq}
\end{equation}
One needs here the wave function $\psi_0(k,r)$ away from the asymptotic region, i.e., for
$0 \leq r \leq \infty$. One can use either the Schr\"odinger equation or the Lippmann-Schwinger
(LS) equation to obtain $\psi_0(k,r)$. Ref.~\cite{Haidenbauer:2018jvl} uses the latter,
the most convenient choice for nonlocal potentials, like those of Refs.~\cite{Vidana:2019amb,
Haidenbauer:2017dua}. Let $T_0(q,k;E)$ denote the $S-$wave component of the half-off-shell 
T-matrix and $\tilde\psi_0(k,r) = \exp(-2i\delta_0)\,\psi_0(k,r)$, where 
$\delta_0 = \delta_0(k)$ is the phase shift; then~\cite{Haftel:1970zz,Joachain}
\begin{eqnarray}
\tilde\psi_0(k,r) &=& j_0(kr) \nn \\
&+& \frac{1}{\pi} \int dq \,q^2 \, j_0(qr)\, \nn \\
&\times&\frac{1}{E-E_1(q)-E_2(q)+i\epsilon} \, T_0(q,k;E) \ ,
\nn \\
\label{Eq:wfb}
\end{eqnarray}
where $E=E_1(k)+E_2(k)$, with $E_i = \sqrt{k^2 + M^2_i}$. The normalization of $\psi_0(k,r)$ is 
\begin{eqnarray}
\psi_0(k,r) \xrightarrow{r\rightarrow\infty} &&
\frac{e^{-i\delta_0}}{kr} \sin(kr + \delta_0) \nn \\
&& = \frac{1}{2ikr}\left[e^{ikr} - e^{-2i\delta_0} e^{-ikr}\right],   
\end{eqnarray}
which differs from the most common form by an overall phase $e^{-2i\delta_0}$, an immaterial
difference as one needs absolute squares only. 
In the case of $\Lc  N$ there are two $S$-waves, namely the $^1S_0$ state with total 
spin $S=0$ and the $^3S_1$ with $S=1$. 
Moreover, the latter partial wave can couple to the $^3D_1$ state via the tensor force. 
In the present study the coupling $^3S_1$-$^3D_1$ is taken into account when solving 
the LS equation and evaluating the corresponding T-matrices $T_{ll'}$ ($l,l'=0,2$),
see, e.g., Ref. \cite{Polinder:2006zh}. However, in the actual calculation of the 
wave function according to Eq.~(\ref{Eq:wfb}), only the $S-$wave component 
$T_{00}$ is needed \cite{Haidenbauer:2018jvl}. 
Standard experiments allow one to measure only an average over the $S=0$ and $1$ states. 
It is commonly assumed that the weight is the same as for free scattering 
which suggests the substitution 
$|\psi_0|^2 \to 1/4\, |\psi_{^1S_0}|^2 + 3/4\,|\psi_{^3S_1}|^2$. 

In the present study we adopt the usual approximation for the source function $S_{12}(r)$ 
and represent it by a Gaussian distribution which depends only on one parameter, namely
the source radius $R$. It is given by $S_{12}(\bold{r})=\exp(-r^2/4R^2)/(2\sqrt{\pi}R)^3$
in the proper normalization. 
In the presence of the Coulomb interaction, i.e. for $\Lambda_c p$, Eq.~(\ref{Eq:wfa}) 
takes on the form \cite{Morita:2016} 
\begin{equation}
\psi(\br,\bk) = \Psi^C(\br,\bk) + \psi^{SC}_0(r,k) - F_0(kr)/(kr) \ ,   
\label{Eq:wfCo}
\end{equation}
where $F_l(kr)$ is the regular Coulomb wave function for $l=0$  and 
$\psi^{SC}_0(r,k)$ the strong scattering wave function in the 
presence of the Coulomb interaction. $\Psi^C(\br,\bk)$ is the full Coulomb
wave function. With these quantities the correlation function $C(k)$ can
be obtained again from Eq.~(\ref{Eq:cq}) after an appropriate substitution of 
the wave functions. Most importantly, one has to keep in mind that the ``1''
in Eq.~(\ref{Eq:cq}) 
has to be replaced by $\int dr\, r^2 S_{12}(r) \int \frac{d\Omega}{4\pi} |\Psi^C(\br,\bk)|^2$
\cite{Morita:2016}. 
How calculations with the Coulomb interaction can be performed in momentum space 
is described in detail in Appendix D of Ref.~\cite{Holzenkamp:1989}. 
For that the Vincent-Phatak method \cite{VP} is employed. With it the 
Coulomb-distorted strong T-matrix can be obtained, on- and half-off shell, 
by a matching condition. Then the scattering wave function $\psi^{SC}_0(r,k)$ can be
again evaluated analogous to Eq.~(\ref{Eq:wfb}).

% % % % % % % % % % % % % % % % % % % % % % % % % % % % % % % % % % % % % % % % % 

\section{Interactions and results}
\label{sec:res}

In this section we present our predictions for 
$\Lc N$ interactions~\cite{Haidenbauer:2020uci,Haidenbauer:2017dua}
obtained by extrapolating lattice simulations of the HAL QCD Collaboration
to the physical point  (LQCD-e) .
We begin with summarizing the main ingredients of the LQCD-e potential. 
Then, we show results for the $\Lc p$ correlation functions obtained from that 
potential and study their source size dependence. 
In addition, we explore the sensitivity of the correlation functions to 
the strength of the $\Lc N$ interaction. For that purpose we resort to 
results of phenomenological potentials available in the literature
\cite{Maeda:2015hxa,Vidana:2019amb,Garcilazo:2019ryw} for orientation. 
As already mentioned, in general these models suggest a more strongly attractive 
$\Lc N$ force than lattice QCD and some \cite{Maeda:2015hxa} even lead to 
two-body bound states.     
It is of interest to examine the impact of such properties on the correlation function. 

\subsection{The $\Lc N$-$\Sc N$ interaction}

The $\Lc N$-$\Sc N$ potential is constructed in close analogy to  the $\La N$-$\Si N$ 
interaction developed by the J\"ulich-Bonn-Munich group \cite{Polinder:2006zh,
Haidenbauer:2013oca,Haidenbauer:2019boi} based on chiral EFT and contains contact terms 
and contributions from one-pion exchange. For the $^1S_0$ and $^3S_1$-$^3D_1$ partial waves 
of interest here, one has~\cite{Haidenbauer:2017dua}:  
\begin{eqnarray}
&&V_{\Lc N}(^1\!S_0) = {\tilde{C}}_{^1\!S_0} + {C}_{^1\!S_0}\, ({p}^2+{p}'^2) \ , \label{1S0} \\
&&V_{\Lc N}(^3\!S_1) = {\tilde{C}}_{^3\!S_1} + {C}_{^3\!S_1}\, ({p}^2+{p}'^2) \ , \label{3S1} \\
&&V_{\Lc N}(^3\!D_1 -\, ^3\!S_1) = {C}_{\varepsilon_1}\, {p'}^2 \ , \label{mix1} \\
&&V_{\Lc N}(^3\!S_1 -\, ^3\!D_1) = {C}_{\varepsilon_1}\, {p}^2 \, , \label{mix2} \\
&&V^{OPE}_{{Y_c N\to Y_c N}} =-f_{{Y_c Y_c\pi}}f_{{NN\pi}}
\frac{\left({\bf \sigma}_1\cdot {\bf q} \right)
\left({\bf \sigma}_2\cdot {\bf q} \right)}{ {\bf q}^{\,2}+m_{\pi}^2} \ .
\label{OPE}
\end{eqnarray}
where $p = |{\bf p}\,|$ and ${p}' = |{\bf p}\,'|$  are the initial and final 
center-of-mass (c.m.) momenta, and ${\bf q} = {\bf p'} - {\bf p}$ the transferred
momentum. 
The strength parameters of the contact terms, ${\tilde{C}}_{i}$ and ${C}_{i}$, 
the so-called low-energy constants (LECs), have been determined in 
Ref.~\cite{Haidenbauer:2017dua} by considering the HAL QCD results for the $^1S_0$ 
and $^3S_1$ phase shifts at unphysical quark masses corresponding to 
$m_\pi=410$~MeV and $570$~MeV 
and by a subsequent extrapolation of the established potential to the physical point, 
guided by chiral EFT. The actual values of the LECs can be found in Table 1 of
that work\footnote{Note that the values for 
$C_{^1S_0}$ and $\tilde C_{^3S_1}$ are erroneously interchanged in Table I of 
Ref.~\cite{Haidenbauer:2017dua}. E.g., $C_{^1S_0} =0.2377\cdot 10^{4}$ GeV$^{-4}$ 
while $\tilde C_{^3S_1} =-0.02077\cdot 10^{4}$ GeV$^{-2}$ for $m_\pi = 138$ MeV, etc.
}.
The coupling constants for pion exchange are given by the $f_{BB'\pi} = g^{BB'}/2\,F_\pi$, 
the ratio of the axial-vector strength $g_A^{BB'}$ 
to the pion decay constant $F_\pi$. For the latter and for $g_A^{NN}$ the 
standard values \cite{PDG} ($F_\pi \approx 93$~MeV, $g_A^{NN}=1.27$) are used 
while the others are fixed from available lattice QCD results close to the physical point, 
amounting to $g_A^{\Sc\Sc}=0.71$~\cite{Alexandrou:2016} 
and $g_A^{\Lc\Sc}=0.74$~\cite{Albertus:2005,Can:2016}. 
Note that, under the assumption of isospin conservation, $f_{\Lc\Lc\pi} \equiv 0$. 
Thus, there is no direct contribution from pion exchange to the $\Lc N$ potential at 
leading order \cite{Polinder:2006zh}. However, it contributes to the $\Lc N$ interaction 
of Ref.~\cite{Haidenbauer:2017dua} via the channel coupling $\Lc N$-$\Sc N$. 

\begin{table*}[tp]
\renewcommand{\arraystretch}{1.5}
\vskip 0.2cm
\caption{Results for effective range parameters of the $\Lambda_c N$ 
and $Y_cN$-A potentials inferred from LQCD and for the simulations
of the potentials from Refs. \cite{Maeda:2015hxa} (CTNN-d), 
\cite{Vidana:2019amb} (Model A), and \cite{Garcilazo:2019ryw} (CQM). 
For the latter the results of the original interactions are given
in brackets. 
}
\begin{center}
\begin{tabular}{l||c|c|c|c}
\hline
\hline
Potential & \ $a_s$ (fm) \ & \ $r_s$ (fm) \ & \ $a_t$ (fm) \ & \ $r_t$ (fm) \ \\
\hline
LQCD-e (500) \cite{Haidenbauer:2017dua} & -0.85 &  2.88 & -0.81  & 3.50 \\
LQCD-e (600) \cite{Haidenbauer:2017dua} & -1.01 &  2.61 & -0.98  & 3.15 \\
LQCD-e (500) \cite{Haidenbauer:2020uci} & -0.85 &  2.88 & -0.79  & 3.58 \\
LQCD-e (600) \cite{Haidenbauer:2020uci} & -1.01 &  2.61 & -0.91  & 3.34 \\
\hline 
CQM  \cite{Garcilazo:2019ryw}    & -0.87 (-0.86) &  4.55 (5.64) & -2.31 (-2.31)  & 2.81 (2.97) \\
Model A \cite{Vidana:2019amb}   & -2.60 (-2.60) &  2.67 (2.86) &-15.88 (-15.87)  & 1.64 (1.64) \\
CTNN-d \cite{Maeda:2015hxa}      &  5.31 (5.31) &  1.20 (1.20) &  5.01 (5.01)  & 1.20 (1.20) \\
\hline
\hline
\end{tabular}
\end{center}
\label{ERE}
\end{table*}
\renewcommand{\arraystretch}{1.0} 

In view of additional lattice results published by the HAL QCD Collaboration recently,
the $\Lc N$ interaction has been revisited in Ref. \cite{Haidenbauer:2020uci}.
The new aspect concerns information on the interaction in the $\Sc N$ channel 
\cite{Miyamoto:2017ynx}, specifically in the $^3S_1$ partial wave. 
It turned out that including a direct $\Sc N$ interaction into the coupled-channel
calculation has only a minor effect on the predicted $\Lc N$ phase shifts at the 
energies of interest here \cite{Haidenbauer:2020uci}. 
Nonetheless, for completeness, we study the effect on the $\Lc p$ 
correlation functions too.

With the interactions defined and the parameters fixed, the next step is to solve the 
LS~equation for the quantity $T_0(q,k;E)$~\cite{Haidenbauer:2017dua}. With it 
one can reconstruct the $\Lc N$ wave functions, utilizing Eq.~(\ref{Eq:wfb}), and then, 
in turn, compute the $\Lc N$ correlation functions. The LS equation requires 
regularization~\cite{Epelbaum:2008ga,Machleidt:2011zz} for the 
potential of Eqs.~(\ref{3S1})-(\ref{OPE}). In Ref.~\cite{Haidenbauer:2017dua}
a cutoff scheme with the regularization function 
$f(p',p) = \exp\left[-\left(p'^4+p^4\right)/\Lambda^4\right]$ 
is used ~\cite{Polinder:2006zh,Haidenbauer:2013oca} ,
with $\Lambda$ values $500$~MeV and $600$~MeV. The choice of the $\Lambda$ 
values is motivated by NLO studies of the $\La N$ and $\Si N$ systems in 
Refs.~\cite{Haidenbauer:2013oca,Haidenbauer:2019boi}. The variations of the
results with $\Lambda$ can be assessed from the bands in the figures below.

As said above, we want explore also in how far differences in the interaction
strength as predicted by other $\Lc N$ potentials are reflected in the pertinent
correlation functions. This goal can be achieved in a simple and efficient way 
within our formalism. We employ the same representation for the $Y_c N$ force 
as for LQCD-e interaction, see Eqs. (\ref{3S1})-(\ref{OPE}), but now we adjust 
the contact terms to the effective range parameters from the models 
by Maeda et al.~\cite{Maeda:2015hxa}, Vida\~na et al.~\cite{Vidana:2019amb}, 
and Garcilazo et al.~\cite{Garcilazo:2019ryw}. This allows us to capture 
the essential features and differences such as the overall strength of the interaction
and the relative strength of the singlet and triplet $S-$waves, 
and, thus, enables us to see the impact of these properties on the correlation 
functions. We want to emphasize that we do not need (and we do not aim at) 
an exact and quantitative reproduction of the results by those potentials
for that purpose. 

\begin{figure}[!htbp]
\vskip -0.5cm 
\includegraphics[scale=0.34]{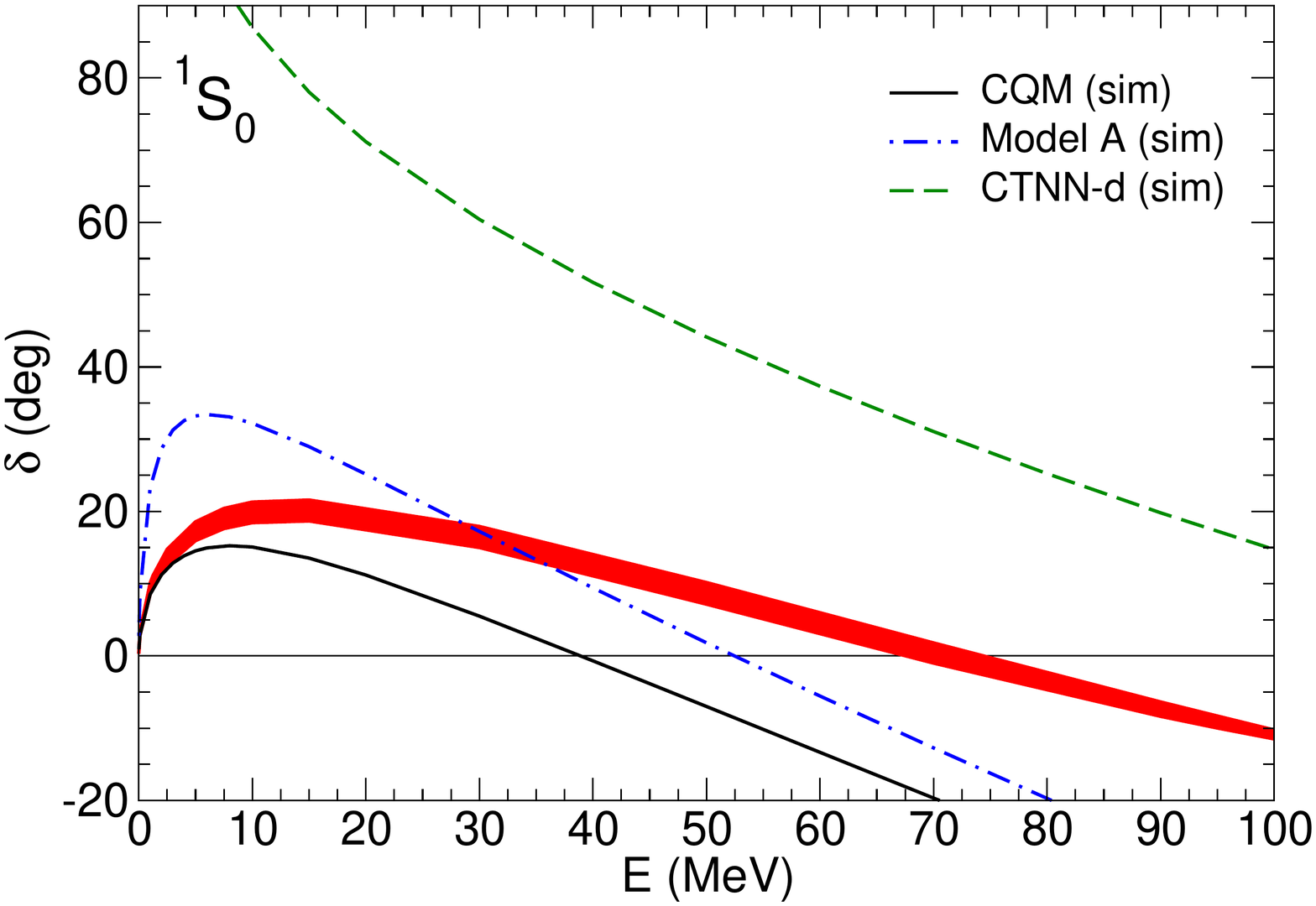}
\includegraphics[scale=0.34]{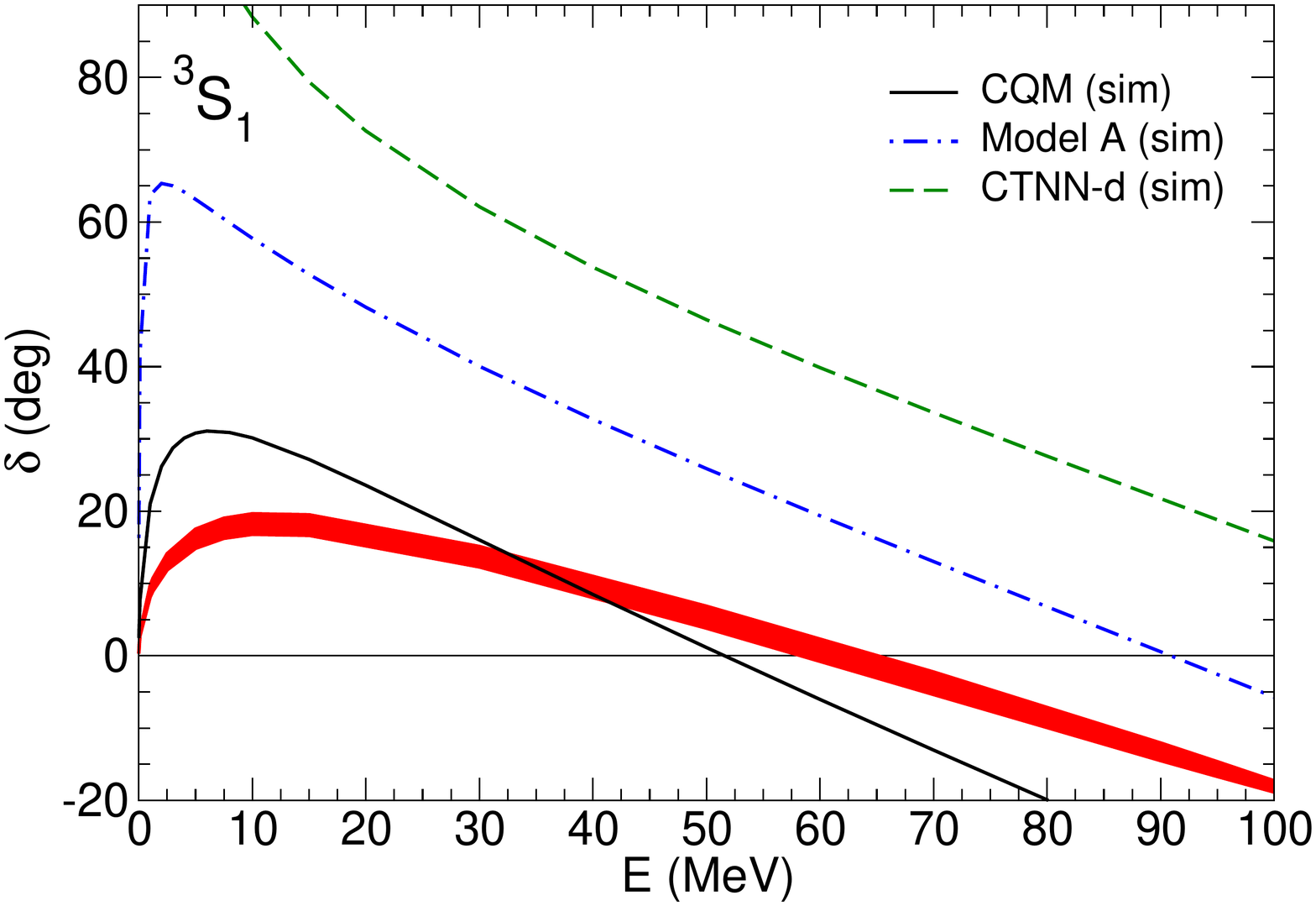}
\caption{$\Lc N$ phases for the $Y_c N$ 
potential inferred from LQCD \cite{Haidenbauer:2017dua} .
The bands represent the cutoff variation $\Lambda = 500-600$ MeV,
see text. In addition results for the 
simulations of the potentials from Refs. \cite{Maeda:2015hxa} (CTNN-d), 
\cite{Vidana:2019amb} (Model A), and \cite{Garcilazo:2019ryw} (CQM)
are shown. 
}
\label{phsim}
\end{figure}

A summary of the $\Lc N$ results is given in Table~\ref{ERE} and in Fig.~\ref{phsim}.
Table \ref{ERE} provides an overview of the $\Lc N$ scattering lengths $a$ and 
effective range parameters $r$ for the various interactions. The first two entries 
are for the LQCD-e interaction from Ref.~\cite{Haidenbauer:2017dua} with cutoffs 
$\Lambda=500,\ 600$ MeV. Then corresponding results for the variant 
considered in Ref. \cite{Haidenbauer:2020uci} ($Y_c N$-A) are listed, 
which includes a direct $\Sc N$ interaction. Finally, one can find results for 
the effective range parameters for our simulations of a selection of models from 
Refs.~\cite{Maeda:2015hxa,Vidana:2019amb,Garcilazo:2019ryw}, together with the
original results in brackets. 
Results for the $^1S_0$ and $^3S_1$ phase shifts are presented in Fig. \ref{phsim}.
From that figure one can read off the different properties immediately. 
It is obvious that the potential from S. Maeda et al.~\cite{Maeda:2015hxa} (dashed lines),
denoted by CTNN-d, is by far the most attractive one. It predicts bound states, as mentioned
before, with binding energies of the order of that of the deuteron in both $S$-waves. 
Model A (dash-dotted lines) presented in the paper by Vida\~na et al.~\cite{Vidana:2019amb}, 
deduced from a $YN$ meson-exchange potential of the J\"ulich group \cite{Reuber:1993ip}  
via SU(4) symmetry arguments, suggests a strongly attractive $^3S_1$ interaction 
and a moderately attractive $^1S_0$ partial wave. 
The $\Lc N$ interaction derived within the constituent-quark model (CQM) by
Garcilazo et al.~\cite{Garcilazo:2019ryw} (solid lines) is closest to the interaction inferred 
from the lattice simulations. Actually, it is slightly less attractive in the  $^1S_0$ 
state but noticeably more attractive in the $^3S_1$ partial wave. 

We note that the results presented above are all obtained without inclusion of the Coulomb
force. Adding the Coulomb interaction leads to a small modification of the effective
range parameters in case of weakly attractive hadron forces like the LQCD-e interactions
\cite{Haidenbauer:2017dua,Haidenbauer:2020uci}.
For example, the singlet (triplet) scattering lengths change
from $-1.01$ fm ($-0.98$ fm) to 
% (500: -0.83,  2.88; -0.80, 3.45)  (600, -0.97, 2.60; -0.96, 3.09) 
$-0.97$ fm ($-0.96$ fm) when Coulomb is added to the LQCD-e (600) potential from 
Ref.~\cite{Haidenbauer:2017dua}. 
There are more sizable effects for strongly attractive potentials like CTNN-d. 
Nonetheless, the bound states survive despite of the Coulomb repulsion, in 
the original model \cite{Maeda:2015hxa} and likewise in our simulation.
% a = 16.0 and 13.6 fm, respectively. 

%%%%%%%%%%%%%%%%
\subsection{Results for the $\Lc p$ correlation function}

\begin{figure}[!htbp]
\vskip -0.5cm 
\includegraphics[scale=0.34]{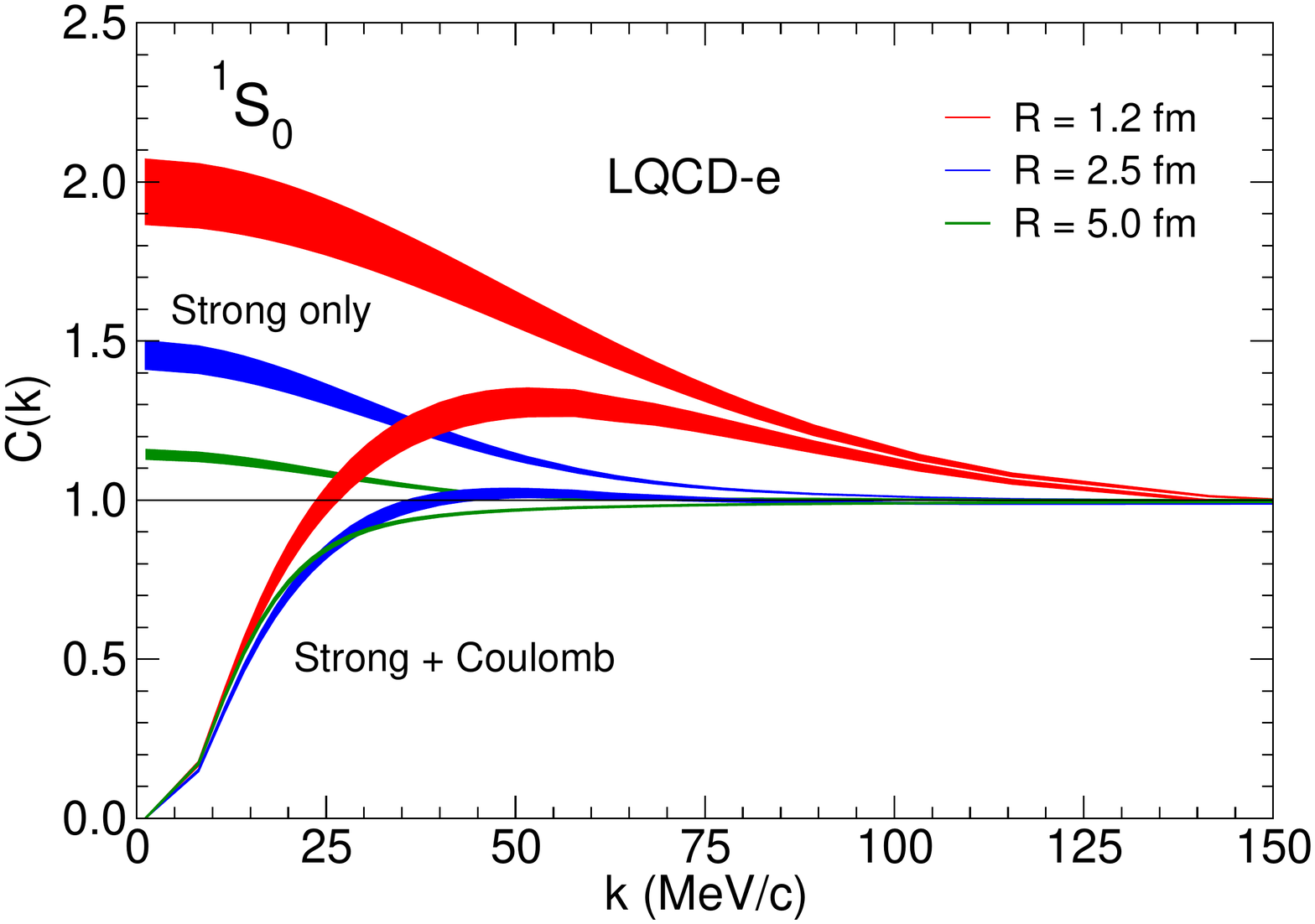}
\includegraphics[scale=0.34]{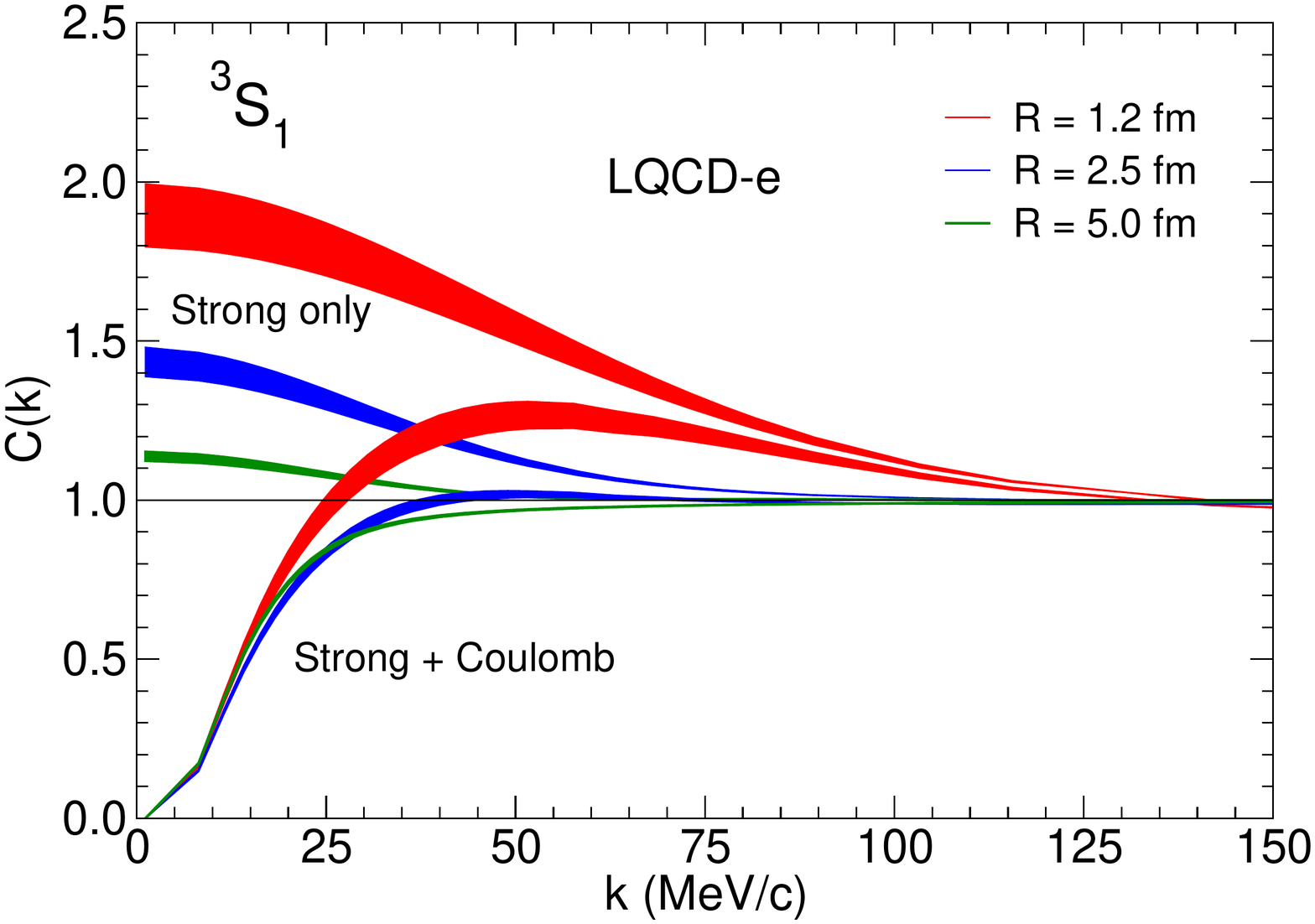} 
\caption{Effect of the Coulomb force and the source size $R$ on the $\Lc p$ correlation
function. The LQCD-e potential \cite{Haidenbauer:2017dua} is used for the calculation. 
}
\label{fig:Lp1} 
\end{figure}

In the discussion of the correlation function 
we start with assessing the effects of the Coulomb interaction and of the source size.
Corresponding results can be found in Fig.~\ref{fig:Lp1}, based on the LQCD-e potential
from Ref.~\cite{Haidenbauer:2017dua}, where we show the $\Lc p$ correlation functions for 
the $^1S_0$ and $^3S_1$ partial waves separately. The choice of considered radii $R$ of 
the Gaussian source is motivated by those suggested in corresponding measurements of
$\La p$ correlation functions in $pp$ collisions at $7$ TeV by the ALICE Collaboration
($R\approx 1.2$ fm) \cite{Acharya:2018gyz} 
and of $\Omega p$ in central and peripheral Au+Au collisions at $200$ GeV
by the STAR Collaboration ($R\approx 2.5,\,5$ fm) \cite{STAR:2018uho}. 

The presence of a repulsive Coulomb force in the $\Lc p$ system leads to a strong
depletion of the correlation function for small momenta.
This effect is well-known and also well-documented, e.g. in calculations and precise 
measurements of $pp$ correlations \cite{Acharya:2018gyz}.
However, since the $\Lc p$ interaction is much less attractive than $pp$, the depletion 
due to Coulomb is noticeable already at larger momenta and it also shifts the maximum in 
the correlation function to somewhat larger momenta. As a consequence the signal due
to the strong interaction is significantly reduced.  Nonetheless, at least for $pp$ collisions
with source radii around $1.2$ fm the effect by the $\Lc p$ interaction should be still 
detectable in an experiment. For heavy-ion collisions with a typical source radius around
$3-5$ fm \cite{Adams:2005,STAR:2018uho} it looks more challenging. 

Comparing the results for $^1S_0$ (top) and $^3S_1$ (bottom) one can see that they
are basically identical for the LQCD-e interaction (without and with Coulomb force). 
This is not too surprising given that the corresponding scattering lengths and 
phase shifts are also almost identical, see Table \ref{ERE} and Fig. \ref{phsim}.
 
\begin{figure}[!tp]
\vskip -0.5cm 
\includegraphics[scale=0.34]{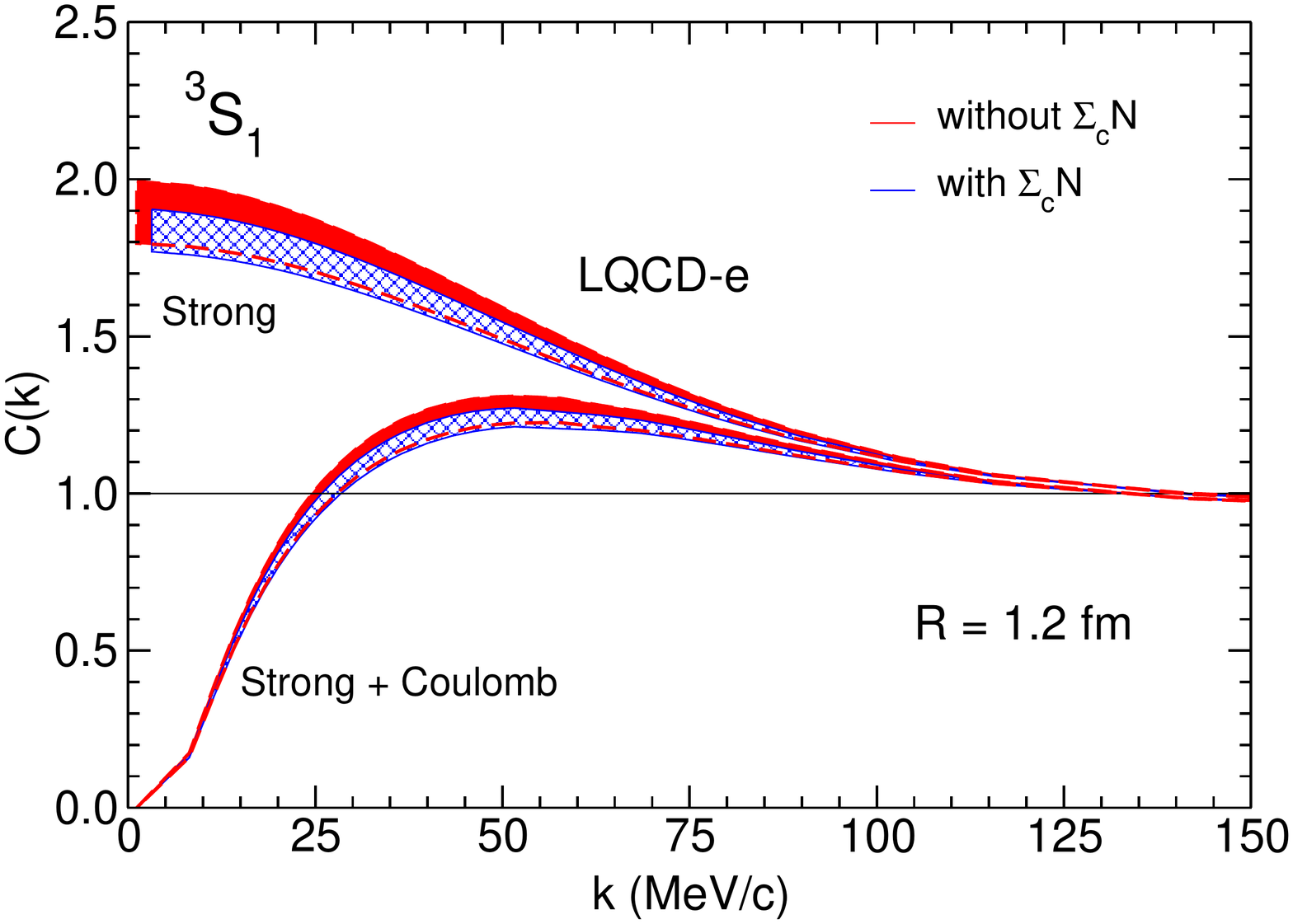}
\includegraphics[scale=0.34]{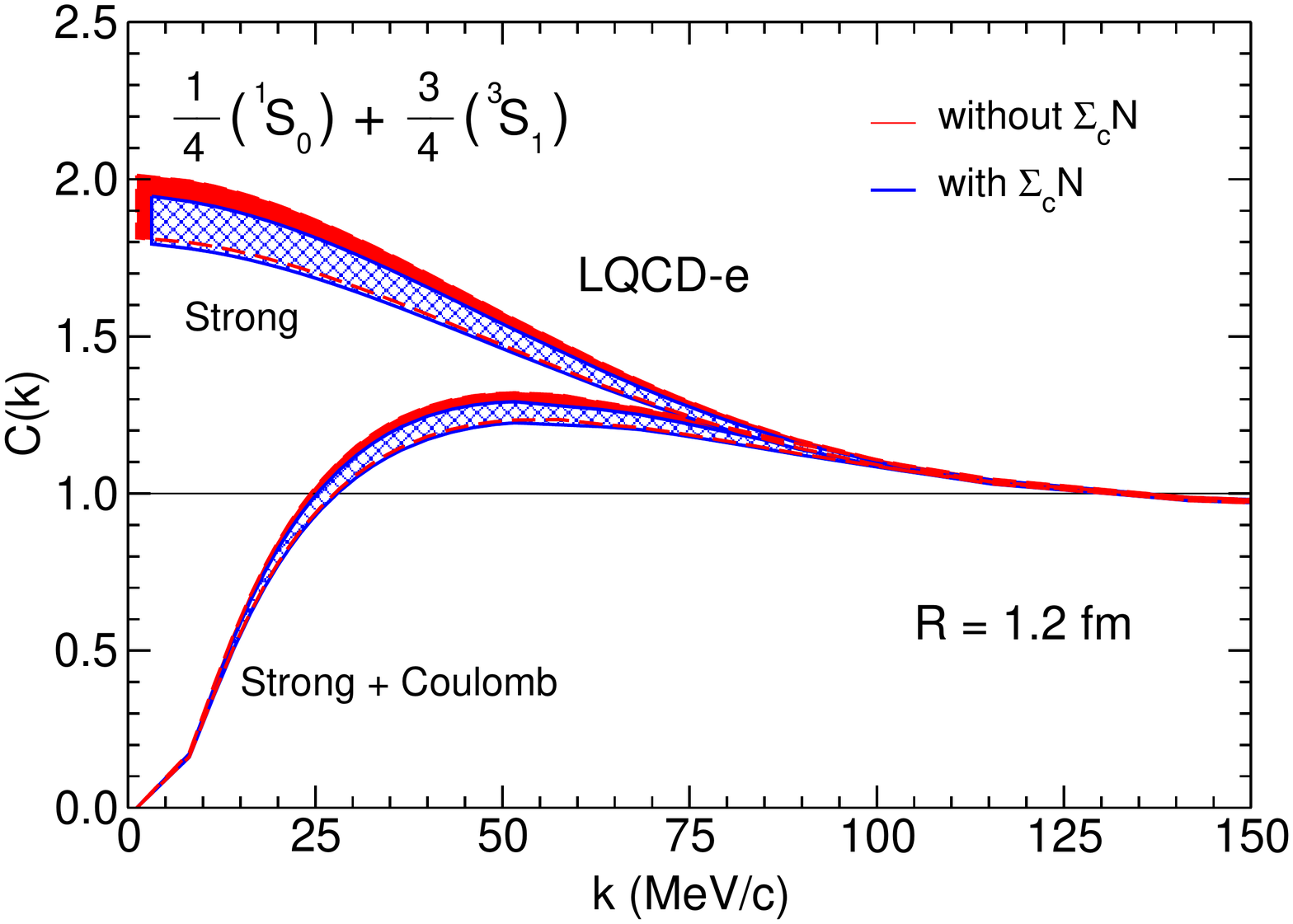}
\caption{Difference in the $\Lambda_c p$ correlation function for the $\Lc N$ 
interactions without \cite{Haidenbauer:2017dua} (filled band) 
and with a direct $\Sc N$ interaction \cite{Haidenbauer:2020uci} (hatched band).
Results are shown for the $^3S_1$ (top) and the spin average (bottom). 
}
\label{fig:Lp3} 
\end{figure}

Next we compare the results for the $\Lc N$ interactions without \cite{Haidenbauer:2017dua}
and with a direct $\Sc N$ interaction \cite{Haidenbauer:2020uci}. This is done in 
Fig. \ref{fig:Lp3}, selectively for the source radius $R=1.2$ fm. 
One can see that there is not much difference. Practically speaking, only the overall 
uncertainy, represented by the band due to the cutoff variation, is somewhat increased 
when  additionally the influence of a direct $\Sc N$ interaction is explicitly taken into 
account.  Therefore, in the following we will show only the results for the potential 
from Ref. \cite{Haidenbauer:2017dua}. 

\begin{figure}[!tp]
\vskip -0.5cm 
\includegraphics[scale=0.34]{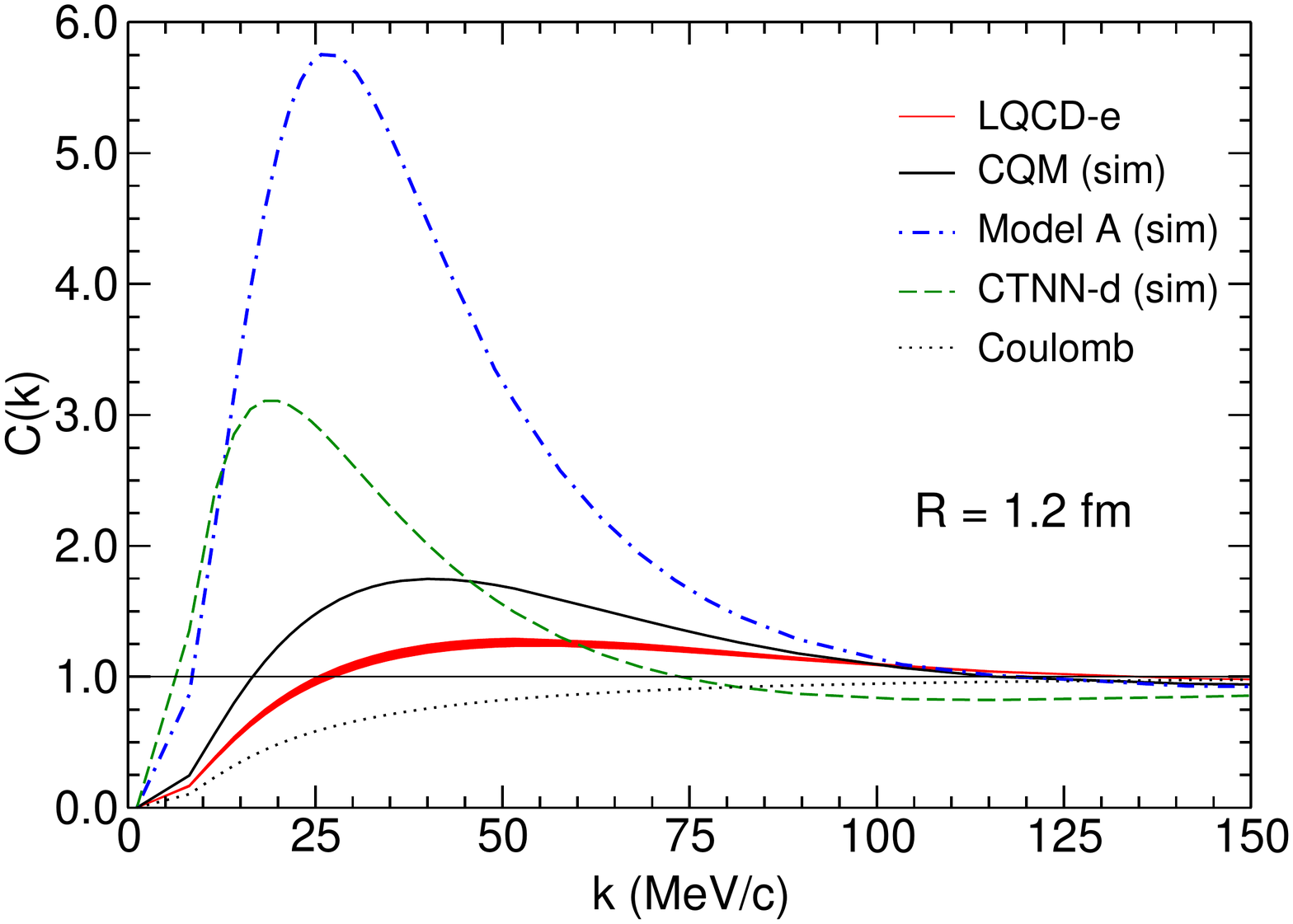}
\includegraphics[scale=0.34]{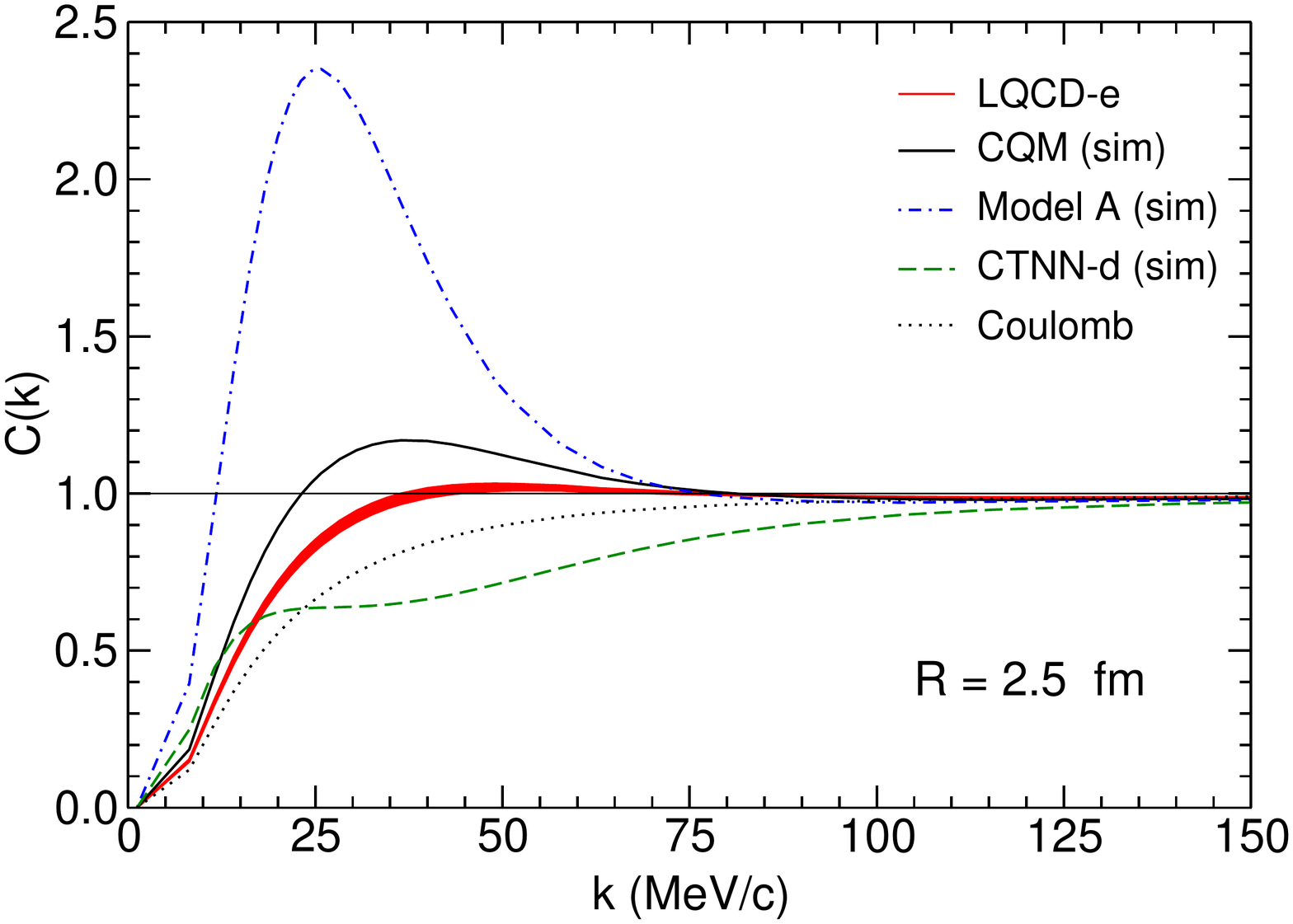}
\includegraphics[scale=0.34]{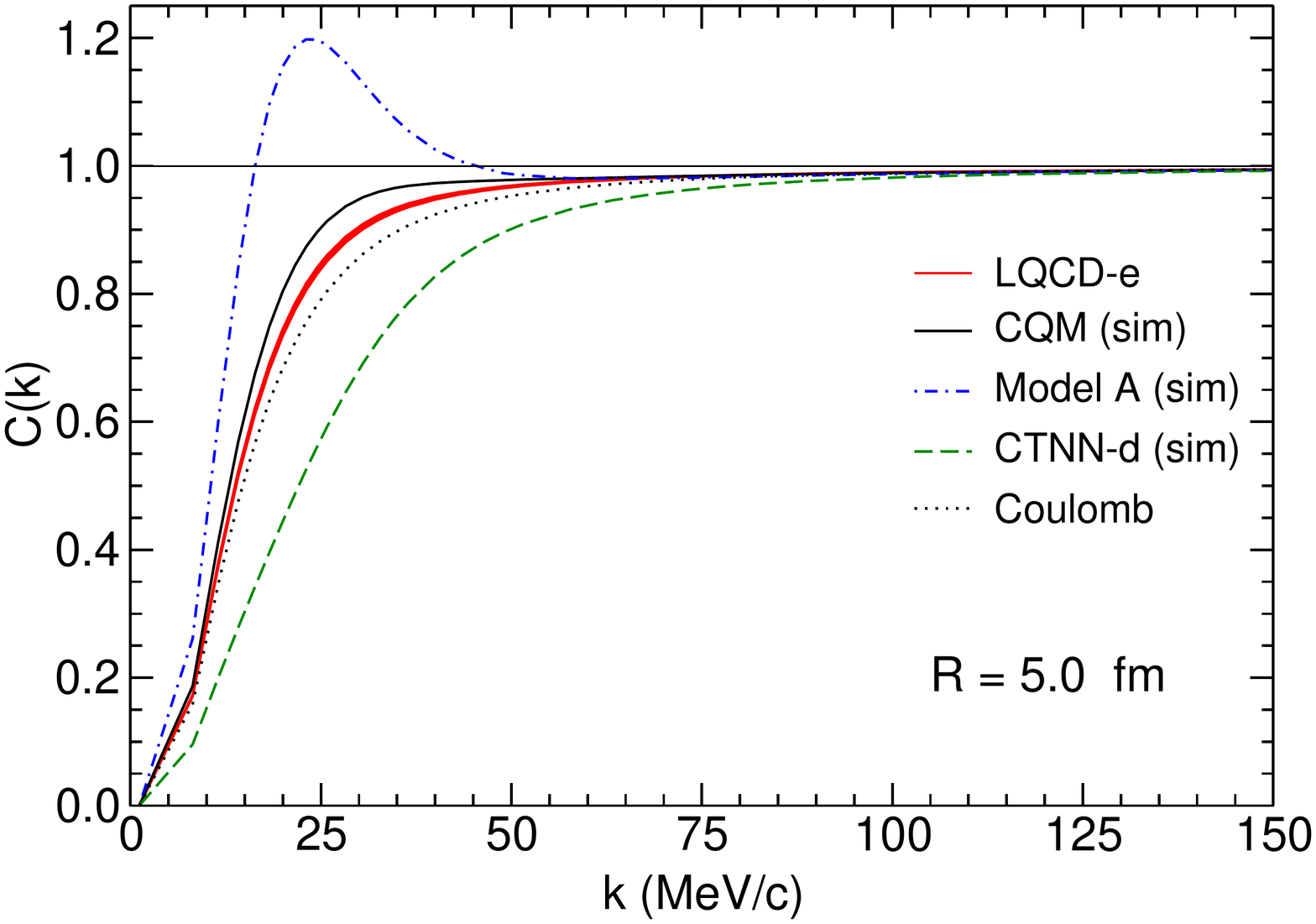}
\caption{
Spin-averaged $\Lc p$ correlation functions including the Coulomb 
interaction for three different source radii $R$. Predictions are shown for the 
LQCD-e interaction (band) and the simulations of CQM \cite{Garcilazo:2019ryw} (solid line),
Model A \cite{Vidana:2019amb} (dash-dotted line), CTNN-d \cite{Maeda:2015hxa} (dashed line).
Also shown is the pure Coulomb interaction (dotted line).
}
\label{fig:Lp2}
\end{figure}

Finally, we contrast the correlation functions predicted by the $\Lc N$ potential 
\cite{Haidenbauer:2017dua} inferred from lattice results with those from (simulated) 
phenomenological potentials. Here we take the spin average in order to be 
as close as possible to the experimental situation. Corresponding results are 
presented in Fig.~\ref{fig:Lp2}, again for different source sizes. 
Already at first sight it is clear that the different potentials
considered lead to quite different predictions for the $\Lc p$ correlation functions. 
Specifically, in general, more attractive interactions yield also larger correlation
functions. Even the simulated CQM interaction which is only moderately more attractive 
than the LQCD-e interaction (cf. the $\Lc N$ phase shifts) yields a noticeably
larger maximum of $C(k)$. This is not least due to the differences in the $^3S_1$
interaction which enters with a three-times larger weight than the $^1S_0$. 
The decisive role of the $^3S_1$ contribution is most prominently seen by 
the result for the simulated model A from Ref. \cite{Vidana:2019amb}, cf. 
dash-dotted lines in Fig. \ref{fig:Lp2}. 
The corresponding correlation function is significantly larger than those of the other
considered interactions and it is still sizable for the source size $R=5$ fm. It is
safe to say that even an experiment with moderate statistics should be sufficient
to discriminate between that model and the properties exhibited by potentials like
CQM or those inferred from lattice simulations (LQCD-e). 
Indeed, given that the spin dependence is not resolved in the standard 
measurements of correlation functions, it is primarily the strength of the 
spin-triplet component which can be tested, of course, always 
under the premises that the actual spin distribution of the produced baryons 
is close to the purely statistical value. 

An interesting behavior is shown by the predictions based on the simulated CTNN-d
interaction that supports bound states. Here there is a delicate interplay between 
the repulsive Coulomb interaction and the strongly attractive $\Lc N$ potential,
which produces a distinct dependence on the source radius. We believe that this 
characteristic behavior constitutes a rather useful signature that could help
for either confirming or ruling out such bound states in experiments. 

%%%%%%%%%%%%%%%%%%%%%%%%%%%%%%%%%%%%%%%%%%%%%%%%%%%%%%%%%

\section{Summary}

We studied the prospects for deducing constraints on the interaction of charmed baryons 
with nucleons from measurements of two-particle momentum 
correlation functions for $\Lambda_c p$. As a benchmark, 
the correlation functions have been evaluated for $\Lambda_c N$ and 
$\Sigma_c N$ interactions extrapolated from lattice QCD simulations by the HAL QCD
collaboration \cite{Miyamoto:2017,Miyamoto:2017ynx}
at unphysical masses of $m_\pi=410-570$ MeV to the physical point using 
chiral effective field theory as guideline \cite{Haidenbauer:2017dua,Haidenbauer:2020uci}. 
In addition, phenomenological $Y_c N$ models from the literature 
\cite{Maeda:2015hxa,Vidana:2019amb,Garcilazo:2019ryw} have been considered 
in order to explore the sensitivity to the properties of the interaction in detail. 
The repulsive Coulomb interaction between the positively charged $\Lambda_c$ and 
the proton has been taken into account in the actual calculation. Only with its effect
included a meaningful and realistic estimate of the signal size that could be
expected in experiments can be given. 

Our studies suggest that the $\Lc p$ correlation function is
definitely a useful tool for acquiring information on the $Y_c N$ interaction. 
Even weakly attractive forces such as those suggested by present-day
lattice simulations lead to effects that should be detectable in pertinent experiments.
In case the $\Lc N$ interaction turns out to be more strongly attractive, as 
predicted by some phenomenological models in the literature, then
measurements of the correlation function would certainly allow one 
to discriminate between the different scenarios. 

An open question at the moment is which yields for $\Lc p$ one
can expect in dedicated experiments. Predictions by different models for production
rates at different accelerators and/or energies have been summarized in the review by 
the ExHIC Collaboration \cite{Cho:2017dcy}, see also Ref. \cite{Steinheimer:2016jjk}.
According to the review,  the expected yields for $\Lc N$ could be as large as
those for $\Omega N$. The latter channel has been already measured by the 
STAR \cite{STAR:2018uho} and ALICE \cite{ALICE:2020} Collaborations. Thus, 
looking at the corresponding data and uncertainties might provide us a rough 
clue on what to expect for $\Lc p$.

\vspace{0.5cm}
\noindent
{\bf Acknowledgements:}
Work partially supported by Conselho Nacional de Desenvolvimento Cien\-t\'{\i}\-fi\-co 
e Tec\-no\-l\'o\-gi\-co (CNPq), Grant. nos. 309262/2019-4 and 464898/2014-5 (G.K), 
and Fun\-da\-\c{c}\~ao de Amparo \`a Pesquisa do Estado de S\~ao Paulo (FAPESP), 
Grant No. 2013/01907-0 (G.K.), 
and also by the DFG and the NSFC through funds provided to the Sino-German CRC 110 
``Symmetries and the Emergence of Structure in QCD'' (DFG grant. no. TRR~110).



\end{document}